# A Novel Human-Based Meta-Heuristic Algorithm: Dragon Boat Optimization

Xiang Li[1,2], Long Lan[1,2], Husam Lahza[3], Shaowu Yang[1,2], Shuihua Wang[4,7], Wenjing Yang[1,2,*], Hengzhu Liu[1,*], Yudong Zhang[3,5,6*]

1. College of Computer Science and Technology, National University of Defense Technology, Changsha 410073, P R China
2. Institute for Quantum Information & State Key Laboratory of High Performance Computing, College of Computer Science and Technology, National University of Defense Technology, Changsha 410073, P R China
3. Department of Information Technology, Faculty of Computing and Information Technology, King Abdulaziz University, Jeddah 21589, Saudi Arabia
4. Department of Biological Sciences, Xi'an Jiaotong-Liverpool University, Suzhou, Jiangsu 215123, China
5. School of Computing and Mathematical Sciences, University of Leicester, Leicester, LE1 7RH, UK
6. School of Computer Science and Engineering, Southeast University, Nanjing, Jiangsu 210096, China
7. Department of Mathematical Sciences, University of Liverpool, Liverpool, L69 3BX, UK.

E-mails: lixiang@nudt.edu.cn, long.lan@nudt.edu.cn, hlahza@kau.edu.sa, shaowu.yang@nudt.edu.cn, shuihuawang@ieee.org, wenjing.yang@nudt.edu.cn, hengzhu_liu@263.net, yudongzhang@ieee.org

* Correspondence should be addressed to, Wenjing Yang, Hengzhu Liu, Yudong Zhang

**Abstract: (Aim)** Dragon Boat Racing, a popular aquatic folklore team sport, is traditionally held during the Dragon Boat Festival. Inspired by this event, we propose a novel human-based meta-heuristic algorithm called dragon boat optimization (DBO) in this paper. **(Method)** It models the unique behaviors of each crew member on the dragon boat during the race by introducing social psychology mechanisms (social loafing, social incentive). Throughout this process, the focus is on the interaction and collaboration among the crew members, as well as their decision-making in different situations. During each iteration, DBO implements different state updating strategies. By modelling the crew's behavior and adjusting the state updating strategies, DBO is able to maintain high-performance efficiency. **(Results)** We have tested the DBO algorithm with 29 mathematical optimization problems and 2 structural design problems. **(Conclusion)** The experimental results demonstrate that DBO is competitive with state-of-the-art meta-heuristic algorithms as well as conventional methods.

## 1. Introduction

Meta-heuristic algorithms, inspired by natural phenomena or concepts, are widely adopted due to their simplicity, flexibility, and efficiency [1]. They have demonstrated their remarkable effectiveness in addressing intricate engineering optimization problems and high-dimensional multi-objective mathematical computations [2]. Therefore, it has been applied in many fields such as machine learning, deep learning, neural networks, mathematical programming [3].

At present, there are more than 540 known meta-heuristic algorithms. Despite the diverse range of them, they can generally be categorized into four types: evolutionary-based algorithms (or called evolutionary algorithms), swarm-based algorithms, physics-based algorithms, and human-based algorithms [4].

Evolutionary-based algorithms are inspired by Charles Darwin's theory of natural evolution. Genetic algorithm (GA)

is an optimization algorithm inspired by the principles of natural selection and heredity in biology [5]. The key attributes of this algorithm include the direct manipulation of structural objects, free from the constraints of derivation and functional continuity. GA inherently possesses implicit parallelism, allowing it to excel in global optimization tasks. Furthermore, it employs a probabilistic optimization strategy that adapts the search direction dynamically. Differential evolution (DE) continuously explores and refines the population to search for the optimal solution through an iterative process of trial and error. This algorithm stands out for its simplicity, ease of implementation, and rapid convergence rate, making it a highly effective and widely adopted random optimization algorithm [6]. Evolutionary programming (EP) employs the Gaussian, Cauchy, Lévy, and single-point mutation operators to tackle optimization problems that are difficult to address solve using a single mutation operator. EP combines the advantages of multiple mutation operators, making it an effective tool for addressing optimization challenges. Other noteworthy evolutionary algorithms encompass bio-inspired optimization (BIO) [7], grammatical evolution (GE) [8], genetic programming (GP) [9].

Swarm-based algorithms draw inspiration from the collective intelligence of swarms, simulating the cooperative behaviors of individuals within a group to efficiently explore solution spaces and discover optimal solutions. Bat algorithm (BA) , inspired by the sonar-like behavior of bats in echolocation, adapts its search strategy and direction in response to environmental changes, making it versatile for diverse task requirements [10]. Bee colony optimization (BCO) employs artificial bees as agents to collaborate in addressing complex combinatorial optimization problems [11]. Cuckoo search (CS), inspired by mandatory parasitic behavior of some cuckoo birds, as well as the Lévy flight pattern of other birds and fruit flies, is designed to locate optimal solutions in optimization problems [12]. CS is known for its robustness and global search capabilities, as well as its advantages in parameter settings. And firefly algorithm (FA) is an algorithm based on the flash behavior of fireflies in nature [12]. There are other notable swarm-based algorithms, such as artificial bee colony (ABC) [13], ant colony optimization (ACO) [14], grey wolf optimizer (GWO) [15], whale optimization algorithm (WOA) [16], virus colony search (VCS) [17], and krill herd algorithm (KHA) [18].

Physics-based algorithms leverage physical principles and natural phenomena to guide the optimization process. It emulates physical phenomena and processes found in nature, constructing models based on physical laws and properties in the pursuit of optimal solutions. Big Bang-Big Crunch (BB-BC) algorithm is rooted in the theory of universe evolution, taking into account energy dissipation, disorder, randomness, and the tendency of randomly distributed particles to form ordered structures [19]. It generates random points using the BB-BC method and consolidates them into a single representative point using either the centroid or least-cost method. The objective is to explore a broader range of solutions and improve search efficiency. Gravitational search algorithm (GSA) simulates the interactions between objects in the universe, considering each solution within the problem's solution space as an object [20]. It calculates the gravitational pull of each solution based on the objective function of the problem, guiding the search process through gravitational force to discover optimal solutions. Multi-verse optimizer (MVO) is an algorithm based on the concepts of white hole, black hole, and wormhole in cosmology. It views the solution space of an optimization problem as a collection of multiple latent universes, each representing a set of feasible solutions. In addition, there are other physics-based optimization algorithms, including electro-magnetism optimization (EMO) [21], harmony search (HS)[22], nuclear reaction optimization (NRO) [23], simulated annealing (SA) [24], sine cosine algorithm (SCA) [25].

Human-based algorithms are inspired by human behavior and decision-making processes. It models the causes and actions of humans when faced with problems and making decisions, guiding the optimization process and finding optimal solutions. Unlike other algorithms that aim to mimic nature intelligence, human-based algorithms focus on the intelligent behavior of humans collaborating to achieve specific objectives. Football game inspired algorithm (FGIA) simulates the behavior of players in a football game, emphasizing teamwork to find the best positions for scoring points under the guidance of a coach [26]. It introduces a unique strategy that effectively balances exploration and development. Artificial coronary circulation system (ACCS) simulates the development of coronary arteries (veins) within the human heart to identify optimal solutions [27]. Each capillary is considered a potential candidate, and starting with a randomized initial population, candidate solutions are evaluated using the coronary growth factor. During each iteration, the most promising candidate is selected as the main coronary vessel (artery or vein), while the remaining capillaries serve as search space explorers. The heart determines whether other candidates are approaching or moving away from the main coronary vessels and employs cardiac memory to seek the best solution. Moreover, there are other human-based optimization algorithms, such as artificial cooperative search (ACS) [28], artificial immune system (AIS) [29], artificial immune network (AIN) [30], exchange market algorithm (EMA) [31], passing vehicle search (PVS) [32], social emotional optimization (SEO) [33], sperm motility algorithm (SMA) [34].

Inspired by the dragon boat racing, our team proposed a novel human-based meta-heuristic optimization algorithm called dragon boat optimization (DBO). This algorithm takes into account various factors, including the social psychological and behavioral states of each crew member during the race, the effects of varying water surface conditions on the paddlers, the adaptive adjustments made by the dragon boat's drummer in response to changing dragon boat ranking, and the collaborative dynamics between the drummer and paddlers. The highlights of this paper are as follows.

(i) Analyze the social psychological influences (social loafing and social incentive) that dragon boat crews experience during dragon boat racing from a social psychology perspective, and the behavior patterns that result from these influences.

(ii) Two factors (acceleration factor and attenuation factor) are introduced to characterize the kinematics indexes by analyzing the different states of cooperative work between the drummer and the paddlers.

(iii) The imbalance rate of the paddlers is introduced by analyzing the impact of water surface conditions on the paddlers during paddling and the paddler's attempt to maintain the optimal angle of entry for the paddle.

(iv) Proposition of two distinct strategies for updating the states of paddlers, which are contingent on different dragon boat rankings. These strategies aim to optimize the performance of the dragon boat crew during races.

The remainder of this paper is organized as follows: Section 2 presents the algorithmic principles and operational mechanism of DBO. Section 3 presents the experimental results of our algorithm compared with other algorithms. In Section 4, the experimental results of optimizing classical engineering problems with various algorithms are presented. And Section 5 concludes this paper.

## 2. Dragon Boat Optimization

The origin of the dragon boat racing is related to Qu Yuan, a patriotic poet of the Chu during the Warring States Period (475-221 B.C.). To commemorate the patriotic poet, a grand dragon boat race is held every year during the Dragon Boat Festival in China. Today, China's dragon boat racing has developed from a local activity into a grand sports event. There are more than 100 countries and regions in the world that hold dragon boat races annually. Dragon boat optimization is the modeling of a dragon boat race as a complete optimization process, where the objective is optimized efficiently by applying multiple strategies.

2.1 Crew with different duties

In dragon boat racing, the swift and steady progress of a dragon boat is intricately tied to the effective teamwork of its crew. Specifically, the crew can be classified into three distinct roles based on their various duties: drummer, paddler, and steersman. The positions they occupy on the dragon boat are illustrated in Figure 1.

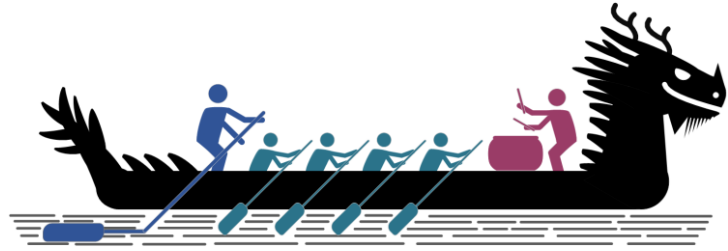

Figure 1 Crew on a dragon boat.

2.1.1 Drummer

The drummer holds the responsibility of regulating the boat's speed by beating the drum. In general, each dragon boat has only one drummer, positioned at the front of the boat, as the rose-red person shown in Figure 1. In dragon boat racing, the drummer should pay attention to the difference in covering distance between their own dragon boat and others. By vigilantly monitoring the distance, the drummer employs the rhythmic drumming to guide the paddlers, signaling them to adapt their paddling pace. This mode can effectively regulate the speed of the dragon boat.

2.1.2 Paddlers

The paddlers are responsible for propelling the dragon boat forward by synchronizing their paddling speed with the rhythm of the drum. Their position, between the drummer and the steersman, is in the middle of the dragon boat, as the peacock-blue person shown in Figure 1. In dragon boat racing, skilled and experienced paddlers excel at mitigating the effects of water surface conditions and hull turbulence, ensuring optimal performance with each stroke.

It's worth emphasizing that achieving the ideal paddle entry angle requires paddlers to maintain a downward gaze and closely observe the water surface conditions throughout the race. At the same time, they must continually adjust their paddling pace based on the rhythm of the drumbeats, thereby fine-tuning the forward momentum of the entire dragon boat.

2.1.3 Steersman

The steersman is responsible for steering the boat and his position is at the stern, as the dark-blue person shown in Figure 1. During the race, the steersman must continually adapt the angle of the tiller to accommodate the ever-changing water conditions. A well-trained steersman can expertly maintain the dragon boat on the correct course, ensuring efficient proceeds along the pre-determined course at all times.

2.2 Optimization algorithm

Inspired by dragon boat racing, the DBO algorithm analyzes the states of the drummers and paddlers on different dragon boats, and deduces the racing states of all of them. Additionally, the algorithm incorporates a social psychology mechanism to model the entire process of a dragon boat race. Figure 2 shows a top view of a dragon boat race.

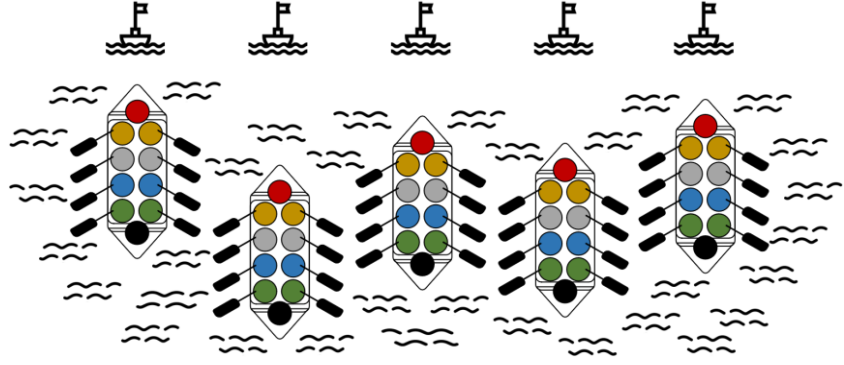

**Figure 2 Top view of a dragon boat race.**

2.2.1 Social behavior patterns

Dragon boat racing is a team sport in which human social attributes play a crucial role. As a result, during the race, the crews are subject to the influence of social psychology mechanisms, namely social loafing and social incentives. Social loafing refers to the phenomenon in which individuals in a group activity exert less effort than they would if working alone, often due to shared responsibilities and reduced individual effort levels. On the other hand, social incentive refers to the utilization of external factors to stimulate enthusiasm, creativity, and effectiveness in individuals or groups when engaging in work or activities, with the aim of improving overall efficiency.

Taking these social psychology mechanisms into account, it can be observed that when applying the DBO to address an unconstrained problem, the dragon boat crew tends to exhibit a social loafing behavior pattern. However, in the case of constrained problems, the constraining conditions serve to boost the motivation of the crew, resulting in a socially incentive behavior pattern.

Therefore, the social behavior factor is introduced to characterize the behavior patterns of the crew. Its formula is as follows.

$$\psi = \begin{cases} \frac{DBN}{R_d}, R_d < DBN \text{ or unconstrained situation} \\ 1, R_d > DBN \text{ or constrained situation} \end{cases} \quad (1)$$

Here, $\psi$ denotes the social behavior factor. $DBN$ denotes the number of dragon boats participating in a dragon boat racing. $R_d$ denotes a random number.

2.2.2 Acceleration factor

The acceleration factor, the rate at which the drum beat changes, is under the control of the drummer. During the dragon boat race, the drummer calculates the rate at which the drum should be beaten. The calculation of this rate takes into consideration several key factors, including the disparity in distance covered between their own dragon boat and other competing boats, as well as the state of the paddlers. To calculate the acceleration factor, the formula is as follows.

$$\lambda = \frac{\psi \times I - 1}{\psi \times I} \quad (2)$$

Here, $\lambda$ denotes the acceleration factor. $I$ denotes the number of iterations.

2.2.3 Attenuation factor

In an ideal scenario, all paddlers should synchronize their paddling with the drumbeat. However, in actual races, each crew member, especially the paddlers, while maintaining a high intensity of exercise, it is difficult to escape the dilemma of strength attenuation. This results in the paddler's performance getting worse as the number of paddling increases. To account for this phenomenon, the attenuation factor is introduced to characterize this state. It is calculated by the following formula.

$$\mu = 1 + \frac{\psi \times I - l}{\psi \times I} \tag{3}$$

Here, $\mu$ denotes the attenuation factor. $l$ denotes the number of iterations when the formula is called.

2.2.4 Imbalance rate of paddlers

The forward propulsion of a dragon boat depends on the paddler's paddling of the paddle, which generates a reaction force on the paddle from the water, propelling the boat forward. To achieve the strongest forward propulsion, the paddler needs to consider several factors before each paddling of the paddle. These factors include the angle, course, and distance of the paddling, as well as the support generated by the paddling and the degree of obstruction when the boat is moving forward. By adhering to the principles of hydrodynamics, paddlers are able to fully utilize properties of the water to generate the strongest driving force.

Well-executed paddling should maintain an optimal angle of entry for the paddle to maximize the propulsion of the dragon boat, even in the face of fluctuations in the water surface. However, when multiple dragon boats are racing parallel to each other, each boat must contend not only with its own inherent water surface fluctuations but also those caused by the paddles of neighboring boats. This gives rise to an imbalanced rate that makes it challenging for paddlers to maintain a consistent paddling state. As illustrated in Figure 3, the optimal angle at which the paddle enters the water can significantly mitigate this issue and enhance overall performance.

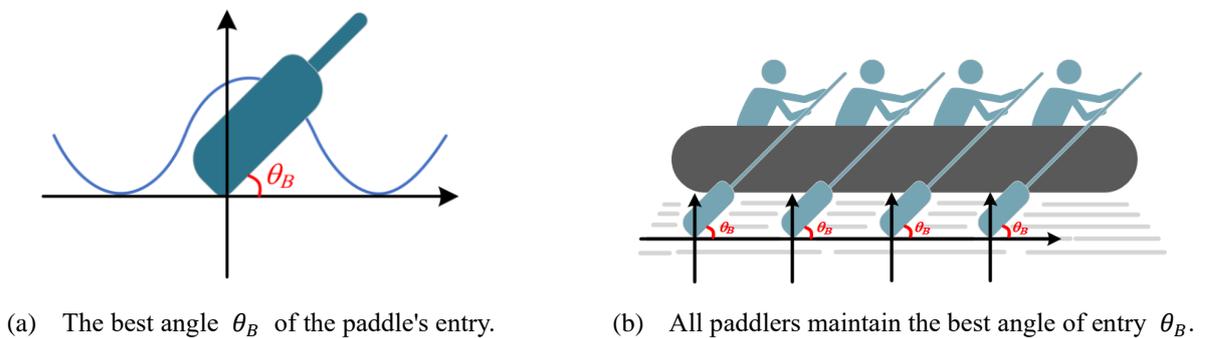

(a) The best angle $\theta_B$ of the paddle's entry.      (b) All paddlers maintain the best angle of entry $\theta_B$.

**Figure 3 The best angle for paddles to entry.**

The best angle for a paddle to entry is $\theta_B$. Here, $\sin\theta$ is employed to calculate the depth of the paddle into the water, while $\cos\theta$ serves as a measure of the resistance of water to the paddle during paddling. Figure 4 shows the imbalance rate curve for different value of $\psi$. The formula for calculating the imbalance rate is as follows.

$$H = \frac{\sqrt{|\cos(\theta)|}}{l*\psi} + H_b \tag{4}$$

Here, $H$ denotes the imbalance rate of a paddler, characterizing the impact of the superposition the fundamental wave from the surface of the water and the wave generated by the progress of other dragon boats on the paddler. $H_b$ denotes the imbalance rate of basis, which is generated from the fundamental wave of the water surface. In this paper, we set the value of $H_b$ to be 0.01. $\theta$ denotes the entry angle of paddle.

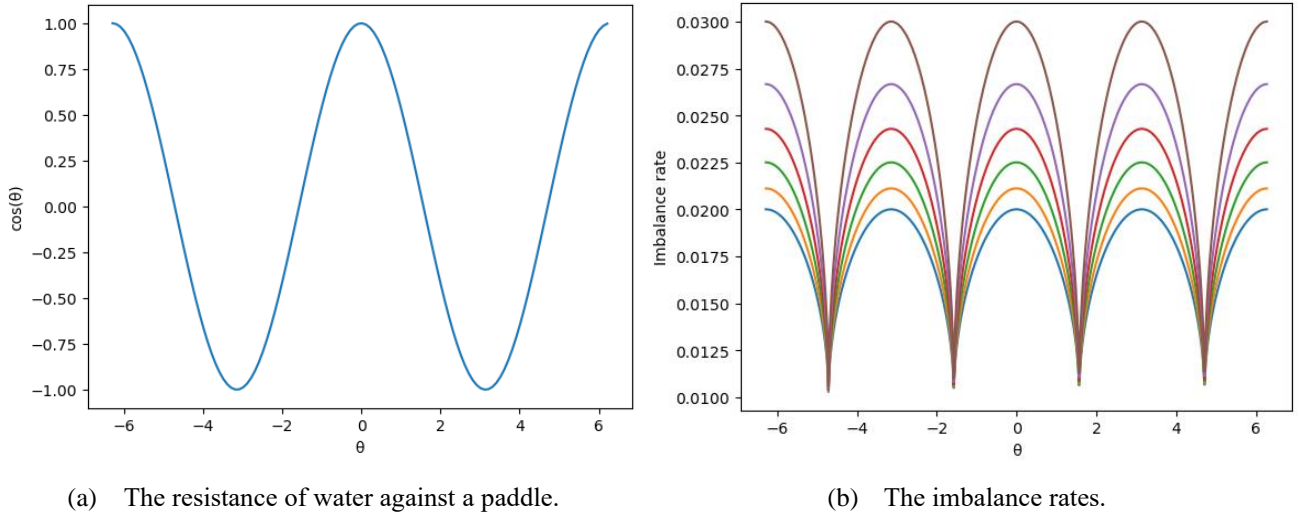

(a) The resistance of water against a paddle.  (b) The imbalance rates.

**Figure 4 An analysis of a paddler's state.**

2.2.5 Strategies for updating crew state

The dragon boat's rapid forward movement relies on the cooperation of the crew. The paddlers' state can be represented as a matrix. When updating these states, the reference object is the state of the paddler in the corresponding position on the fastest dragon boat. However, for the paddlers on the fastest dragon boat, a different update strategy is employed. As a result, the update strategies for paddler states are specifically categorized into two cases: one for the paddlers on the fastest dragon boat and another for those on the other dragon boats. The state update strategies calculate formulas as follows.

$$Paddlers = \begin{bmatrix} g_1^1 & g_2^1 & \cdots & g_{k-1}^1 & g_k^1 \\ g_1^2 & g_2^2 & \cdots & g_{k-1}^2 & g_k^2 \\ \vdots & \vdots & \ddots & \vdots & \vdots \\ g_1^{j-1} & g_2^{j-1} & \cdots & g_{k-1}^{j-1} & g_k^{j-1} \\ g_1^j & g_2^j & \cdots & g_{k-1}^j & g_k^j \end{bmatrix} \tag{5}$$

$$G_f = Paddlers[1,k] = g_k^1, f = 1 \tag{6}$$

$$G_e = Paddlers[j,k] = g_k^j, e \neq 1 \tag{7}$$

$$R_f = G_f \times \lambda \tag{8}$$

$$R_e = \frac{G_f + G_e}{2 \times \mu} \times \lambda \tag{9}$$

Here, $Paddlers$ denotes the state matrix of all paddlers. $G_f$ denotes the state of the paddlers on the fastest dragon boat.

$G_e$ denotes the state of the paddlers on the other dragon boat. $R_f$ denotes the state update strategy for paddlers on the fastest dragon boat. $R_e$ denotes the other state update strategy for paddlers on the other dragon boat.

2.2.6 Computational complexity analysis

The computational complexity of DBO primarily consists of the following parts: data initialization, dragon boat ranking calculation, paddlers states update. Here, when the maximum number of iterations is *I*, the number of dragon boats is *N*, and the number of paddlers (dimension of the problem) is *E*, the computational complexity of dragon boat optimization is *O(N)*. Figure 5 shows the flow chart of DBO. And Table 1 shown the pseudocode of DBO.

**Table 1 The pseudocode of dragon boat optimization.**

| | |
|---|---|
| Step 1 | Initialization parameters: number of iterations, number of dragon boats, number of paddlers on each boat, acceleration factor, attenuation factor, state of paddlers, imbalance rate, elitism parameter. |
| Step 2 | Update the state for each dragon boat |
| Step 3 | while termination conditions: <br> **Step 3.1 Perform elitism** <br> **Step 3.2 Update the states of paddlers** <br> for each dragon boat: <br>    for each paddler: <br>      if the paddler belongs the fastest dragon boat: use $R_f$ to update the state of paddlers <br>      else: use $R_e$ to update the state of paddlers <br> **Step 3.3 Check paddlers' state and update the state for each dragon boat** <br> for each dragon boat: <br>    for each paddler: <br>      if random number < imbalance rate: update the state of paddler by random <br>      else: keep the state of the paddler <br> **Step 3.4 Update the state for each dragon boat** |
| Step 4 | Output of the winning dragon boat |

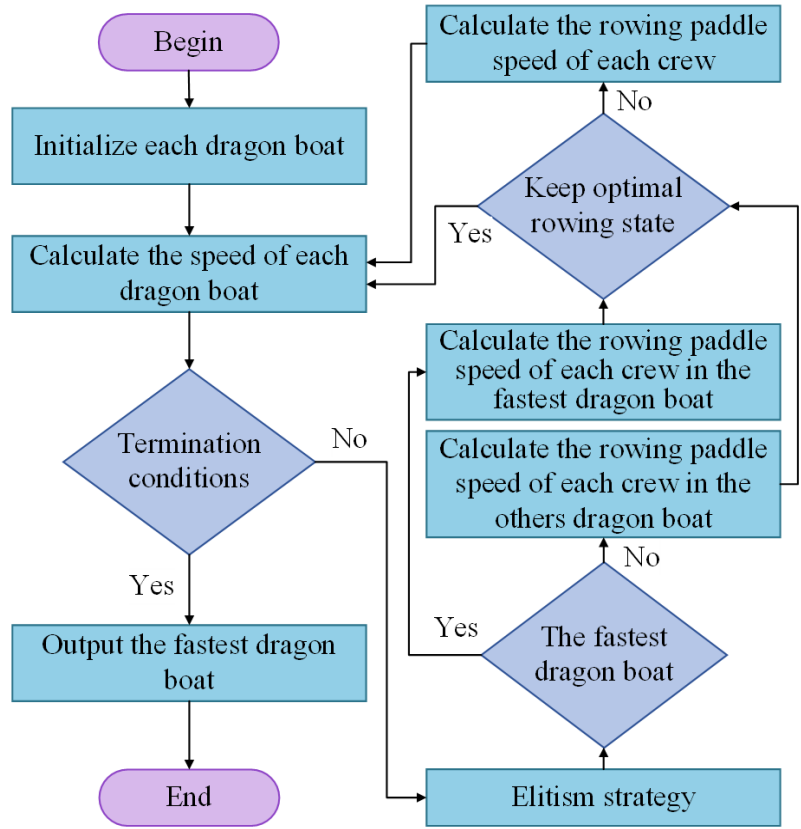

Figure 5 The flow chart of dragon boat optimization.

## 3. Experimental Results

In this section, we test the performance of DBO on 29 benchmark functions and compare it with other meta-heuristic algorithms. To ensure consistency in the evaluation, the same number of iterations and the dimension of the optimization space are set for all algorithms in the experiments. Table 2 shows the unimodal test functions (F1-F7). Figure 6 shows the 3-D versions of unimodal test functions. Table 3 shows the multimodal test functions (F8-F13) used in the experiments. Figure 7 shows the 3-D versions of multimodal test functions (F8-F13). Table 4 shows the fixed-dimension multimodal test functions (F14-F23) used in the experiments. Figure 8 shows the 3-D versions of fixed-dimension multimodal test functions. Table 5 shows the hybrid composition functions (F24-F29). Here, *M* denotes multimodal, *ω* denotes rotated, *N* denotes non-separable, *S* denotes scalable, *D* denotes dimension. Figure 9 shows the 3-D versions of hybrid composition test functions (F24-F29). Table 6 shows the parameter values for the comparative algorithms.

Table 2 Unimodal test functions (F1–F7).

| Function | Description | Dim | Range | $f_{min}$ |
|---|---|---|---|---|
| F1 | $f(x) = \sum_{i=1}^{n} x_i^2$ | 50 | [-100,100] | 0 |
| F2 | $f(x) = \sum_{i=0}^{n}|x_i| + \prod_{i=0}^{n}|x_i|$ | 50 | [-10,10] | 0 |
| F3 | $f(x) = \sum_{i=1}^{d}(\sum_{j=1}^{i} x_j)^2$ | 50 | [-100,100] | 0 |
| F4 | $f(x) = max_i\{|x_i|, 1 \leq i \leq n\}$ | 50 | [-100,100] | 0 |
| F5 | $f(x) = \sum_{i=1}^{n-1}[100(x_i^2 - x_{i+1})^2 + (1 - x_i)^2]$ | 50 | [-30,30] | 0 |
| F6 | $f(x) = \sum_{i=1}^{n}([x_i + 0.5])^2$ | 50 | [-100,100] | 0 |
| F7 | $f(x) = \sum_{i=0}^{n} ix_i^4 + random[0,1)$ | 50 | [-128,128] | 0 |

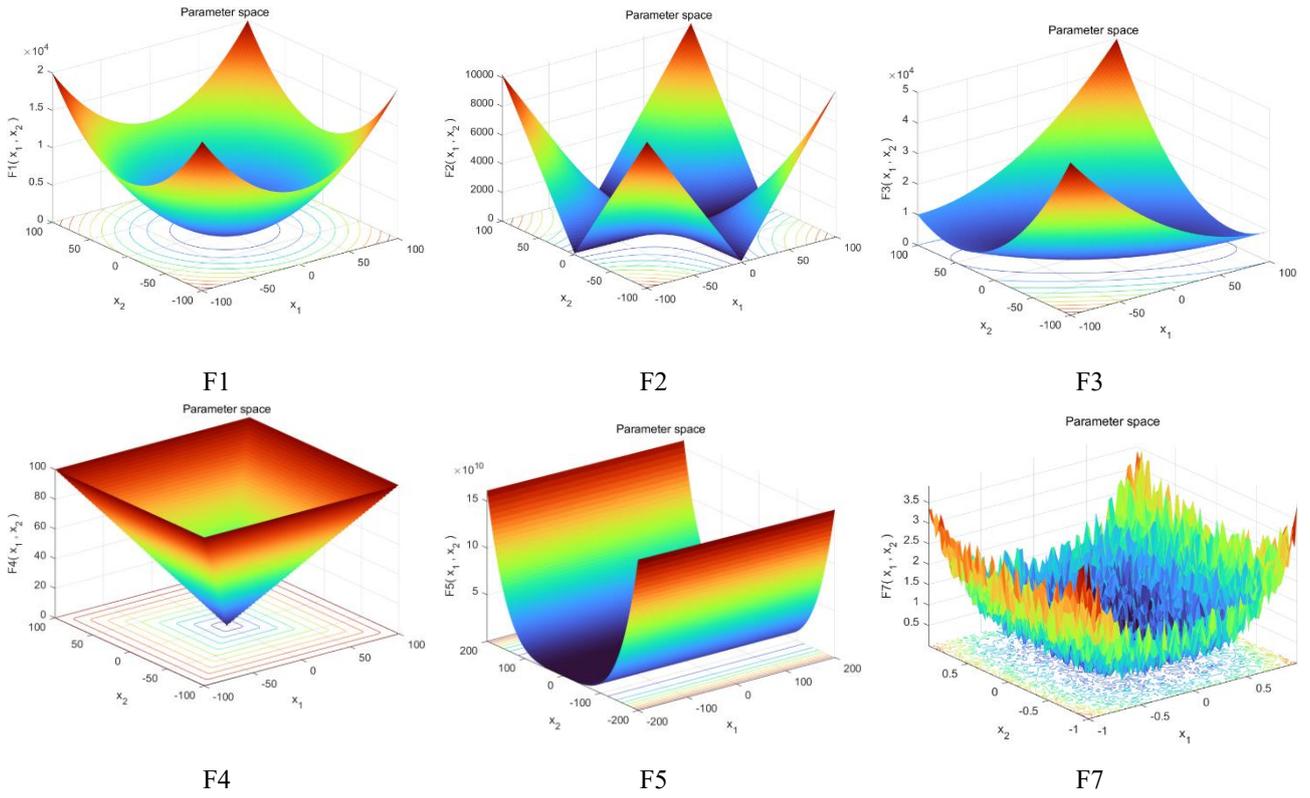

Figure 6 3-D maps for unimodal test functions (F1–F7).

Table 3 Multimodal test functions (F8–F13).

| Function | Description | Dim | Range | $f_{min}$ |
|---|---|---|---|---|
| F8 | $f(x) = \sum_{i=1}^{n}(-x_i \sin(\sqrt{|x_i|}))$ | 50 | [-500,500] | $-418.9829 \times n$ |
| F9 | $f(x) = \sum_{i=1}^{n}[x_i^2 - 10\cos(2\pi x_i) + 10]$ | 50 | [-5.12,5.12] | 0 |
| F10 | $f(x) = -20\exp\left(-0.2\sqrt{\frac{1}{n}\sum_{i=1}^{n}x_i^2}\right) - \exp\left(\frac{1}{n}\sum_{i=1}^{n}\cos(2\pi x_i)\right) + 20 + e$ | 50 | [-32,32] | 0 |
| F11 | $f(x) = 1 + \frac{1}{4000}\sum_{i=1}^{n}x_i^2 - \prod_{i=1}^{n}\cos\left(\frac{x_i}{\sqrt{i}}\right)$ | 50 | [-600,600] | 0 |
| F12 | $f(x) = \frac{\pi}{n}[10\sin(\pi y_1)] + \sum_{i=1}^{n-1}(y_i-1)^2[1 + 1\sin^2(\pi y_{i+1}) + \sum_{i=1}^{n}u(x_i,10,100,4)]$, where $y_i = \frac{x_i+5}{4}$, $u(x_i,a,p,m) = \begin{cases} p(x_i-a)^m & x_i > a \\ 0 & if -a < x_i < a \\ p(-x_i-a)^m & x_i < -a \end{cases}$ | 50 | [-50,50] | 0 |
| F13 | $f(x) = 0.1\{(\sin^2(3\pi x_1) + \sum_{i=1}^{n}(x_i-1)^2[1 + \sin^2(3\pi x_i+1)] + (x_n-1)^2[1 + \sin^2(2\pi x_n)]\} + \sum_{i=1}^{n}u(x_i,5,100,4)$ | 50 | [-50,50] | 0 |

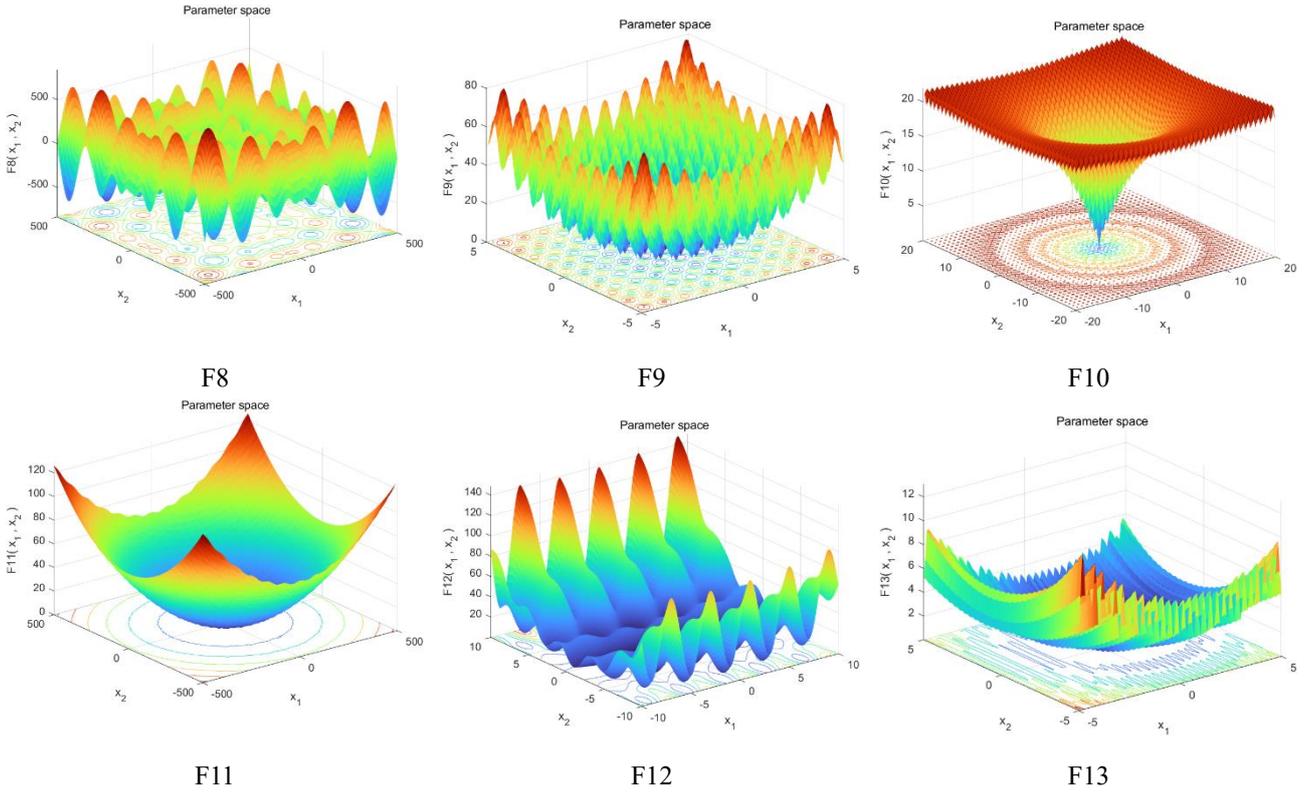

Figure 7 3-D maps for multimodal test functions (F8–F13).

Table 4 Fixed-dimension multimodal test functions (F14-F23).

| Function | Description | Dim | Range | $f_{min}$ |
|---|---|---|---|---|
| F14 | $f(x) = \left(\frac{1}{500} + \sum_{j=1}^{25} \frac{1}{j+\sum_{i=1}^{2}(x_i-a_{ij})^6}\right)^{-1}$ | 2 | [-65,65] | 1 |
| F15 | $f(x) = \sum_{i=1}^{11}[a_i - \frac{x_1(b_i^2+b_i x_2)}{b_i^2+b_i x_3+x_4}]^2$ | 4 | [-5,5] | 0.00030 |
| F16 | $f(x) = 4x_1^2 - 2.1x_1^4 + \frac{1}{3}x_1^6 + x_1 x_2 - 4x_2^2 + 4x_2^4$ | 2 | [-5,5] | -1.0316 |
| F17 | $f(x) = (x_2 - \frac{5.1}{4\pi^2}x_1^2 + \frac{5}{\pi}x_1 - 6)^2 + 10\left(1-\frac{1}{8\pi}\right)cosx_1 + 10$ | 2 | [-5,5] | 0.398 |
| F18 | $f(x) = [1 + (1 + x_1 + x_2)^2(19 - 14x_1 + 3x_1^2 - 14x_2 + 6x_1 x_2 + 3x_2^2)] \times [30 + (2x_1 - 3x_2)^2 \times (18 - 32x_1 + 12x_1^2 + 48x_2 - 36x_1 x_2 + 27x_2^2)]$ | 2 | [-2,2] | 3 |
| F19 | $f(x) = -\sum_{i=1}^{4} c_i \exp\left(-\sum_{i=1}^{3} a_{ij}(x_j - p_{ij})^2\right)$ | 3 | [-1,2] | -3.86 |
| F20 | $f(x) = -\sum_{i=1}^{4} c_i \exp\left(-\sum_{i=1}^{6} a_{ij}(x_j - p_{ij})^2\right)$ | 6 | [0,1] | -3.32 |
| F21 | $f(x) = -\sum_{i=1}^{5}[(X - a_i)(X - a_i)^T + c_i]^{-1}$ | 4 | [0,1] | -10.1532 |
| F22 | $f(x) = -\sum_{i=1}^{7}[(X - a_i)(X - a_i)^T + c_i]^{-1}$ | 4 | [0,1] | -10.4028 |
| F23 | $f(x) = -\sum_{i=1}^{10}[(X - a_i)(X - a_i)^T + c_i]^{-1}$ | 4 | [0,1] | -10.5363 |

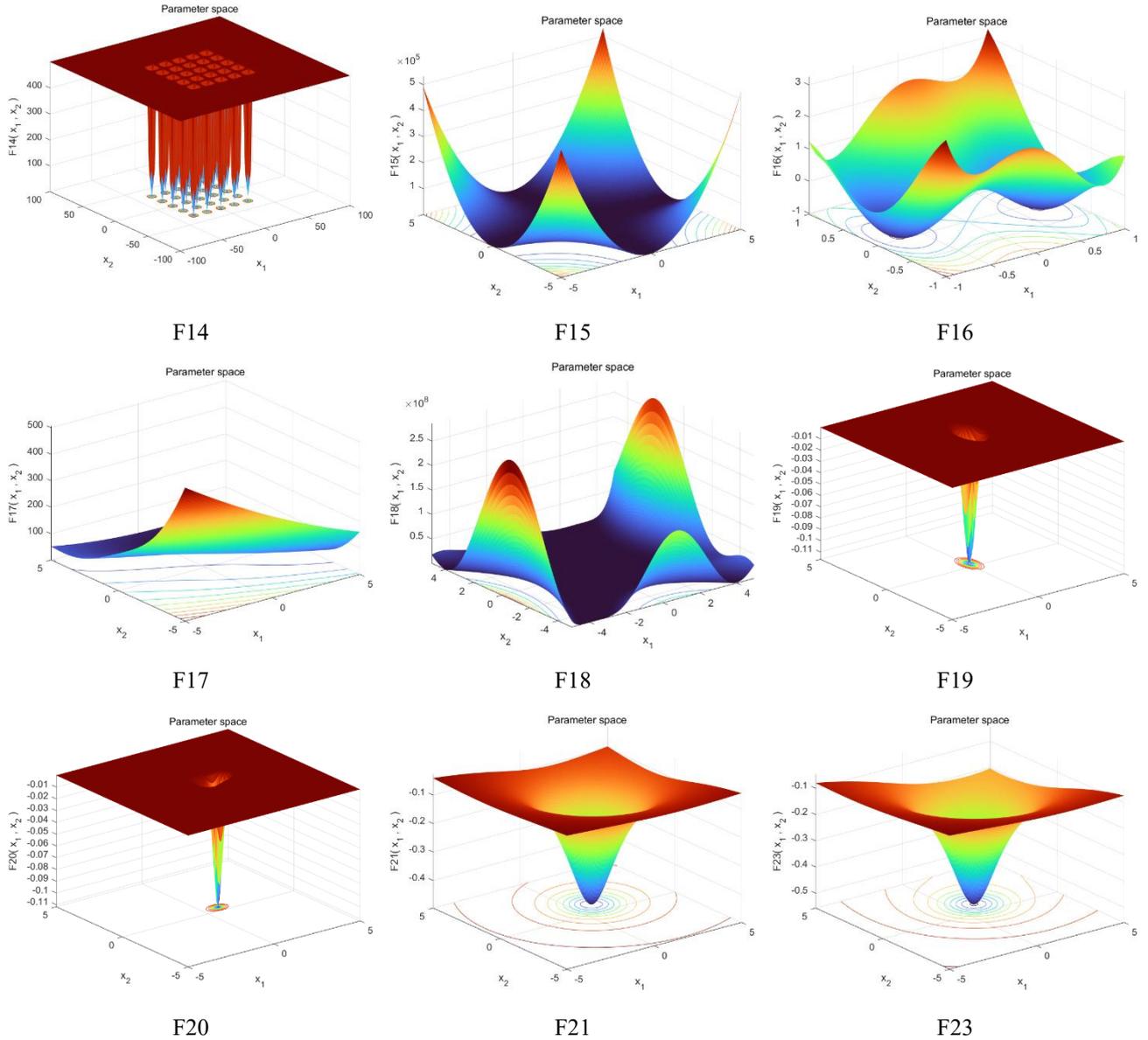

**Figure 8 3-D maps for fixed-dimension multimodal test functions.**

**Table 5 Hybrid composition functions F24-F29.**

| Function | Description | Properties | Dim | Range |
|---|---|---|---|---|
| F24 | $f_{1,2}(x)$: Rastrigin's Function; $f_{3,4}(x)$: Weierstrass Function; $f_{5,6}(x)$: Griewank's Function; $f_{7,8}(x)$: Ackley's Function; $f_{9,10}(x)$: Sphere Function; $\lambda = [1, 1, 10, 10, 5/60, 5/60, 5/32, 5/32, 5/100, 5/100]$; $M_i$ are different linear transformation matrixes with condition number of 2; $\sigma_i = 1$, for $i = 1, 2, \ldots, D$. | M, ω, N, S | 30 | $[-5,5]^D$ |
| F25 | $f_{1,2}(x)$: Ackley's Function; $f_{3,4}(x)$: Rastrigin's Function; $f_{5,6}(x)$: Sphere Function; $f_{7,8}(x)$: Weierstrass Function; $f_{9,10}(x)$: Griewank's Function; $\sigma = [1, 2, 1.5, 1.5, 1, 1, 1.5, 1.5, 2, 2]$; $\lambda = [2*5/32; 5/32; 2*1; 1; 2*5/100; 5/100; 2*10; 10; 2*5/60; 5/60]$; $o_{10} = [0,0,\ldots,0]$; $M_i$ are all rotation matrices. Condition numbers are $[2, 3, 2, 3, 2, 3, 20, 30, 200, 300]$. | M, ω, N, S | 30 | $[-5,5]^D$ |
| F26 | All settings are the same as F25 except $\sigma = [0.1, 2, 1.5, 1.5, 1, 1, 1.5, 1.5, 2, 2]$; $\lambda = [0.1*5/32; 5/32; 2*1; 1; 2*5/100; 5/100; 2*10; 10; 2*5/60; 5/60]$. | $M, \omega, N, S$ | 30 | $[-5,5]^D$ |
| F27 | All settings are the same as F25 except after load the data file, set $o_{1(2j)} = 5$, for $j = 1, 2, \ldots, \lfloor D/2 \rfloor$. | $M, \omega, N, S$ | 30 | $[-5,5]^D$ |
| F28 | $f_{1,2}(x)$: Rotated Expanded Scaffer's F6 Function (CEC05-21); $f_{3,4}(x)$: Rastrigin's Function; $f_{5,6}(x)$: F8F2 Function (CEC05-21); $f_{7,8}(x)$: Weierstrass Function; $f_{9,10}(x)$: Griewank's Function; $\sigma = [1, 1, 1, 1, 1, 2, 2, 2, 2, 2]$; $M_i$ are all orthogonal matrix; $\lambda = [5*5/100; 5/100; 5*1; 1; 5*1; 1; 5*10; 10; 5*5/200; 5/200]$. | $M, \omega, N, S$ | 30 | $[-5,5]^D$ |
| F29 | $f_1(x)$: Weierstrass Function; $f_2(x)$: Rotated Expanded Scaffer's F6 Function (CEC05-21); $f_3(x)$: Rotated Expanded Scaffer's F8F2 Function (CEC05-21); $f_4(x)$: Ackley's Function; $f_5(x)$: Rastrigin's Function; $f_6(x)$: Griewank's Function; $f_7(x)$: Non-Continuous Expanded Scaffer's Function; $f_8(x)$: Non-Continuous Rastrigin's Function; $f_9(x)$: High Conditioned Elliptic Function; $f_{10}(x)$: Sphere Function with Noise in Fitness; $\sigma_i = 1$, for $i = 1, 2, \ldots, D$; $\lambda = [10; 5/20; 1; 5/32; 1; 5/100; 5/50; 1; 5/100; 5/100]$; $M_i$ are all rotation matrices, condition numbers are $[100, 50, 30, 10, 5, 5, 4, 3, 2, 2]$. | $M, \omega, N, S$ | 30 | $[-5,5]^D$ |

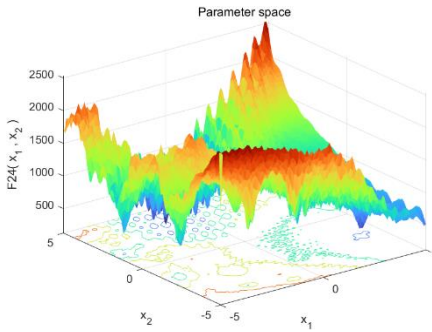

F24

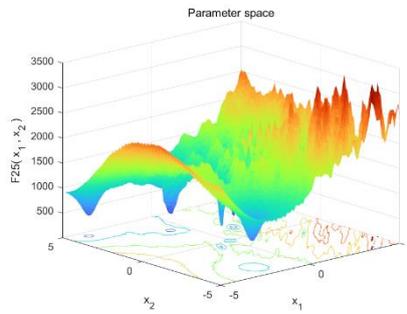

F25

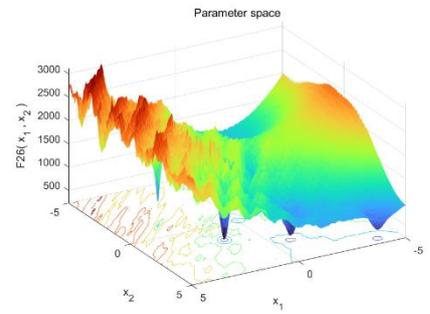

F26

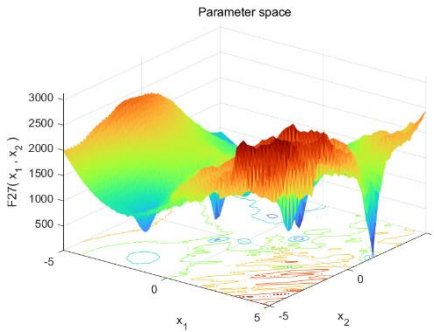

F27

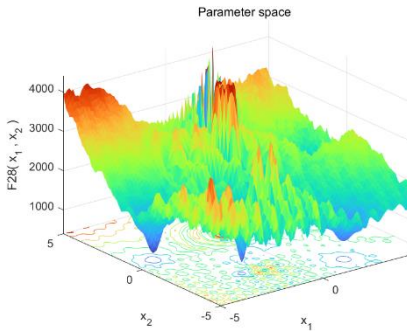

F28

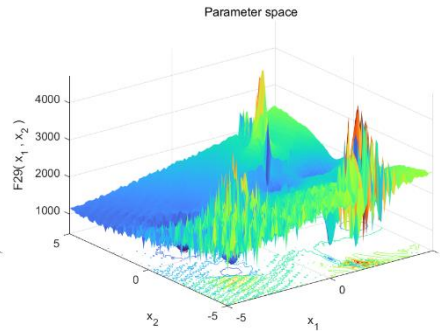

F29

Figure 9 3-D maps for hybrid composition test functions (F24-F29).

Table 6 Parameter values for the comparative algorithms.

| Algorithm | Parameter | Value |
|---|---|---|
| AOA | $\alpha, \mu$ | 5, 0.5 |
| EO | $a_1, a_2$ | 2, 1 |
|  | Generation probability $GP$ | 0.5 |
| GWO | Convergence constant $a$ | [2,0] |
| HHO | Default constant in the levy flight $\beta$ | 1.5 |
| PSO | Inertia weight $w$ | Linear reduction from 0.9 to 0.1 |
|  | Cognitive and social constant $c_1, c_2$ | 2,2 |
| SFA | Thresholds of phase $tv_1, tv_2$ | 0.5, 0.3 |
|  | Initial number of the probability of losing contact $p_0$ | 0.25 |
| WOA | Convergence constant $a$ | [2,0] |
|  | Spiral factor $b$ | 1 |

3.1 Evaluation on unimodal test functions

Table 7 presents the results of unimodal test functions (F1–F7). The experimental results demonstrate that DBO exhibits superior convergence and stability on unimodal test functions (F1-F7). In fact, SFA is second only to DBO in terms of performance on the F5 function. While the convergence value of the SFA for F7 is comparable to that of the DBO, the DBO turns out to be more stable. Figure 10 illustrates the convergence curves of the algorithms for the unimodal test functions.

Table 7 Results of unimodal test functions (F1–F7).

|  |  | AOA | DBO | EO | GWO | HHO | PSO | SFA | WOA |
|---|---|---|---|---|---|---|---|---|---|
| F1 | Ave | 4.6332e-06 | 6.6591e-156 | 1.4245e-21 | 5.3628e-24 | 3.7491e-100 | 3.3913e-02 | 9.2699e-136 | 4.9212e-84 |
|  | Std | 1.6904e-06 | 4.8973e-155 | 1.1831e-21 | 4.1473e-24 | 2.0298e-99 | 2.2509e-02 | 5.0072e-135 | 2.4193e-83 |
| F2 | Ave | 2.1428e-03 | 9.3541e-84 | 1.1636e-13 | 1.3922e-14 | 2.5084e-53 | 2.0265e+01 | 1.9458e-71 | 9.2989e-52 |
|  | Std | 1.7952e-03 | 6.1578e-84 | 6.3903e-14 | 7.3057e-15 | 8.8571e-53 | 1.3177e+01 | 9.2215e-71 | 4.6110e-51 |
| F3 | Ave | 1.1528e-03 | 1.9678e-147 | 2.8218e+01 | 1.1231e-02 | 1.8187e-77 | 8.6051e+02 | 2.2280e-135 | 1.4961e+05 |
|  | Std | 7.8544e-04 | 3.2596e-145 | 3.3447e+01 | 1.7053e-02 | 9.9597e-77 | 2.4240e+02 | 1.2203e-134 | 2.4409e+04 |
| F4 | Ave | 1.6784e-02 | 3.4973e-75 | 2.1256e-02 | 2.8063e-05 | 9.4698e-51 | 2.6892e+00 | 7.8269e-68 | 6.3473e+01 |
|  | Std | 1.1126e-02 | 2.7053e-71 | 2.4533e-02 | 2.1247e-05 | 3.3784e-50 | 4.0120e-01 | 4.2869e-67 | 2.5524e+01 |
| F5 | Ave | 2.7862e+01 | 1.8342e-02 | 4.5147e+01 | 4.7011e+01 | 1.9657e-02 | 2.7650e+02 | 1.6325e-02 | 4.7750e+01 |
|  | Std | 2.2688e-01 | 1.0721e-02 | 4.0261e-01 | 6.9681e-01 | 1.6431e-02 | 1.5732e+02 | 1.6192e-02 | 3.6794e-01 |
| F6 | Ave | 3.1352e+00 | 7.9017e-05 | 2.5053e-02 | 1.9156e+00 | 7.4108e-05 | 3.3678e-02 | 3.8012e-03 | 4.5692e-01 |
|  | Std | 2.2857e-01 | 7.7716e-04 | 7.5512e-02 | 5.3227e-01 | 9.9175e-05 | 2.3635e-02 | 4.0351e-03 | 2.3745e-01 |
| F7 | Ave | 5.6643e-05 | 1.4279e-05 | 4.4231e-03 | 1.9511e-03 | 9.3641e-05 | 2.1925e+01 | 1.6593e-03 | 1.9379e-03 |
|  | Std | 4.5464e-05 | 1.7036e-05 | 1.6139e-03 | 8.5802e-04 | 7.8797e-05 | 1.9564e+01 | 1.4214e-03 | 1.5611e-03 |

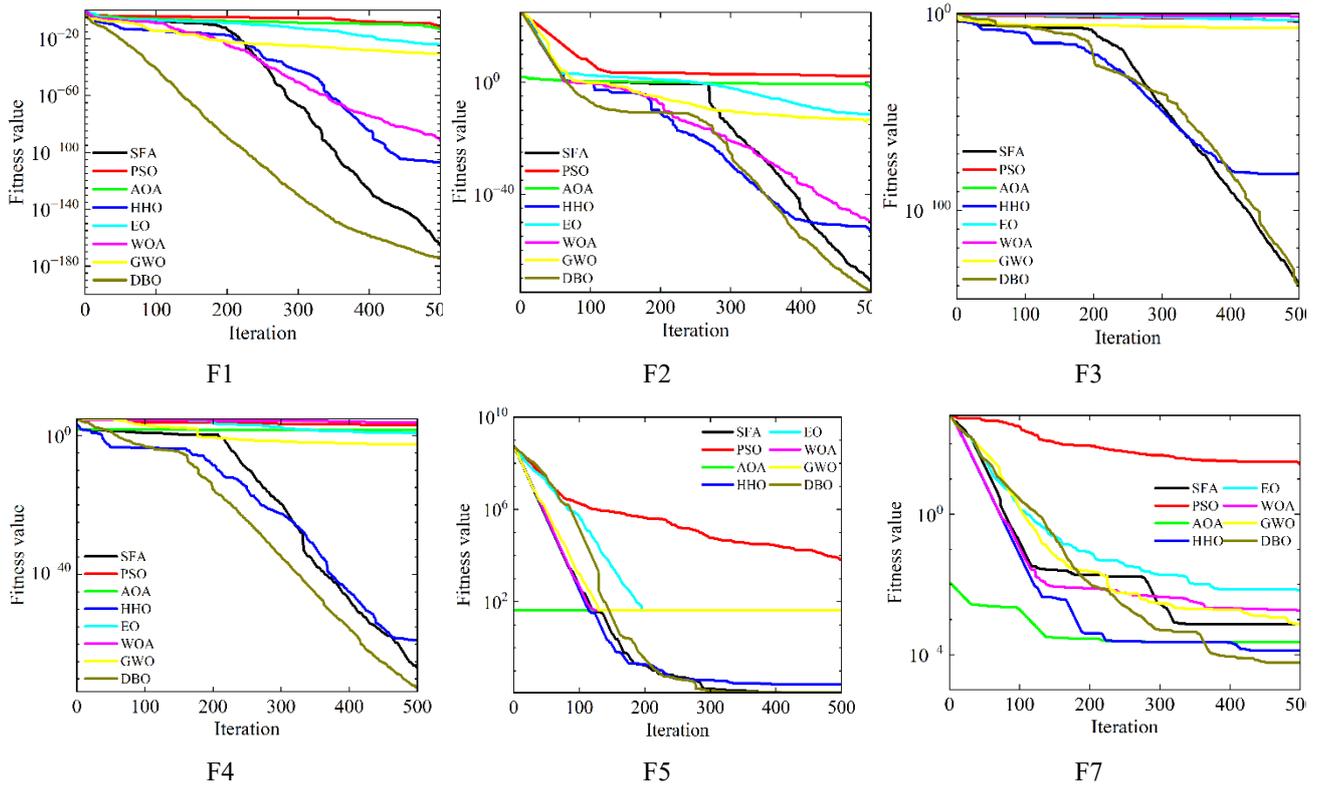

F1  F2  F3

F4  F5  F7

**Figure 10 Convergence curves of the algorithms for the unimodal test functions.**

3.2 Evaluation on multimodal test functions

Table 8 presents the results of multimodal test functions (F8–F13). The experimental results demonstrate that DBO is capable of converging to the minimum value on both test functions F9 and F11. Moreover, it exhibits satisfactory optimization performance in other test functions as well. Figure 11 illustrates the convergence curves of the algorithms for the multimodal test functions (F8–F13).

**Table 8 Results of multimodal test functions (F8–F13).**

|     |     | AOA | DBO | EO | GWO | HHO | PSO | SFA | WOA |
| --- | --- | --- | --- | --- | --- | --- | --- | --- | --- |
| F8  | Ave | -5.4862e+03 | -7.5901e+04 | -1.1632e+04 | -9.0472e+03 | -2.0949e+04 | -8.7464e+03 | -2.0949e+04 | -1.8315e+03 |
|     | Std | 4.3661e+02 | 5.1762e-02 | 7.7830e+02 | 1.4962e+03 | 9.0530e-01 | 2.0282e+03 | 8.8381e-01 | 2.5328e+03 |
| F9  | Ave | 1.5661e-06 | 0.0000e+00 | 8.0839e+00 | 2.4067e+00 | 0.0000e+00 | 2.6546e+02 | 0.0000e+00 | 0.0000e+00 |
|     | Std | 8.7428e-07 | 0.0000e+00 | 3.9042e+00 | 3.5157e+00 | 0.0000e+00 | 3.6313e+01 | 0.0000e+00 | 0.0000e+00 |
| F10 | Ave | 4.4105e-04 | 1.9038e-17 | 6.1170e-12 | 5.4753e-13 | 8.8818e-16 | 9.9041e-01 | 8.8818e-16 | 4.9146e-15 |
|     | Std | 1.9622e-04 | 0.5874e-16 | 3.0387e-12 | 2.6438e-13 | 0.0000e+00 | 5.8354e-01 | 0.0000e+00 | 2.2340e-15 |
| F11 | Ave | 2.3628e-05 | 0.0000e+00 | 5.7536e-04 | 1.7196e-03 | 0.0000e+00 | 5.9138e-03 | 0.0000e+00 | 0.0000e+00 |
|     | Std | 1.1152e-05 | 0.0000e+00 | 2.2018e-03 | 4.6305e-03 | 0.0000e+00 | 7.2094e-03 | 0.0000e+00 | 0.0000e+00 |
| F12 | Ave | 7.4282e-01 | 4.8527e-06 | 1.0538e-03 | 7.2133e-02 | 2.1928e-06 | 2.1428e-02 | 2.0363e-05 | 1.1217e-02 |
|     | Std | 2.7439e-02 | 4.8396e-06 | 2.0283e-03 | 3.1509e-02 | 3.3109e-06 | 4.1309e-02 | 2.8392e-05 | 7.1098e-03 |
| F13 | Ave | 2.9659e+00 | 7.0371e-05 | 6.0907e-02 | 1.5011e+00 | 4.4561e-05 | 5.0377e-02 | 1.8056e-04 | 5.5041e-01 |
|     | Std | 1.2652e-05 | 6.9804e-05 | 6.5702e-02 | 3.6721e-01 | 6.8784e-05 | 3.9608e-02 | 1.6167e-04 | 2.2557e-01 |

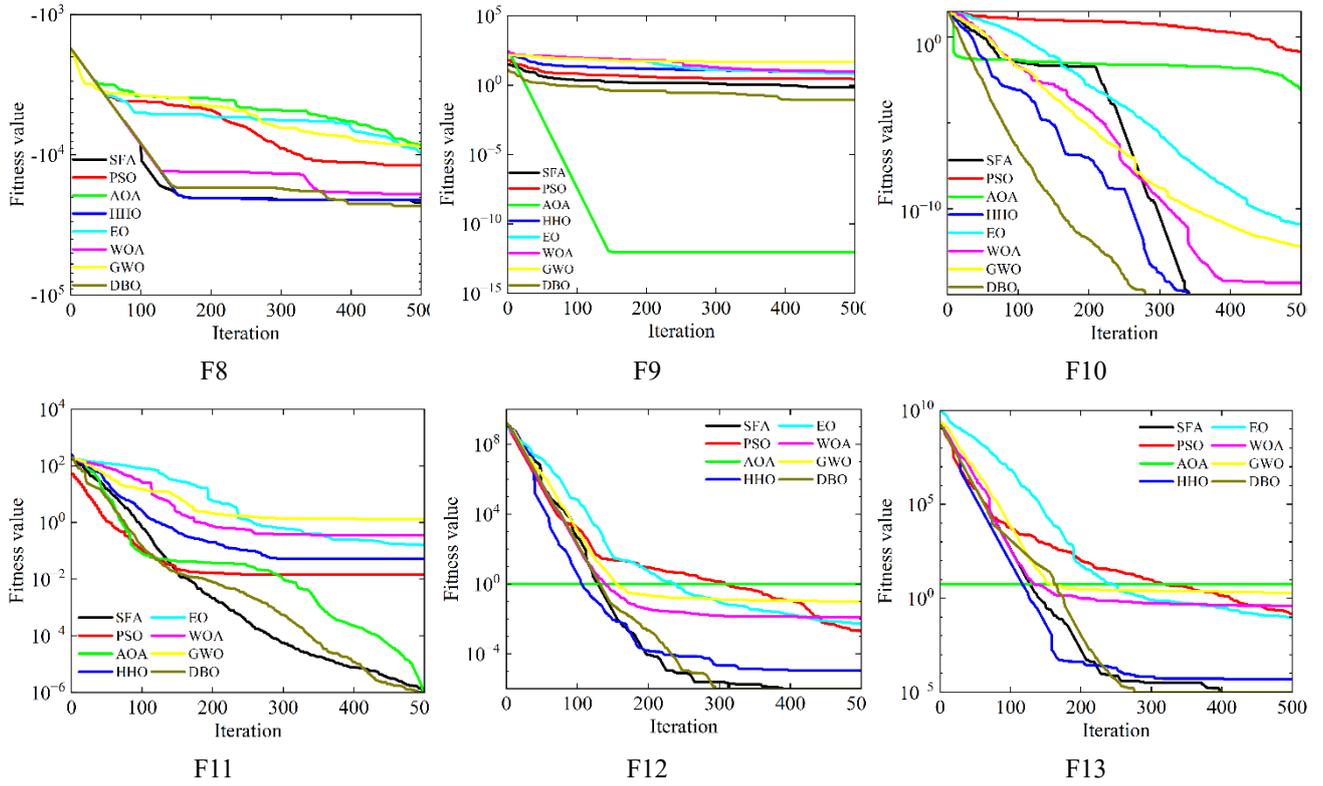

F8 F9 F10
F11 F12 F13

**Figure 11 Convergence curves of the algorithms for the multimodal test functions (F8–F13).**

3.3 Evaluation on fixed-dimension multimodal test functions

Table 9 presents the results of fixed-dimension multimodal test functions (F14-F23). Here, DBO demonstrates excellent performance by converging to the minimum values on the test functions F14, F16-F19, F21 and F23 respectively. Moreover, it also exhibits robustness against various perturbations in the input data. Figure 12 illustrates the convergence curves of the algorithms for the fixed-dimension multimodal test functions.

**Table 9 Results of fixed-dimension multimodal test functions (F14–F23)**

|  |  | AOA | DBO | EO | GWO | HHO | PSO | SFA | WOA |
|---|---|---|---|---|---|---|---|---|---|
| F14 | Ave | 9.1842e+00 | 9.9917e-01 | 9.9802e-01 | 4.5560e+00 | 1.2618e+00 | 1.5930e+00 | 9.9809e-01 | 2.7663e+00 |
|  | Std | 3.9652e+00 | 7.9869e-09 | 1.4857e-16 | 4.2063e+00 | 9.3204e-01 | 9.9137e-01 | 9.9296e-11 | 3.0314e+00 |
| F15 | Ave | 4.4328e-03 | 5.9837e-04 | 2.4089e-03 | 4.3822e-03 | 3.8699e-04 | 5.5717e-03 | 1.2426e-03 | 8.0036e-04 |
|  | Std | 6.9826e-03 | 2.0366e-04 | 6.1573e-03 | 8.2079e-03 | 2.3118e-04 | 8.4903e-03 | 3.3508e-03 | 4.9874e-04 |
| F16 | Ave | -1.3016e+00 | -1.3016e+00 | -1.3016e+00 | -1.3016e+00 | -1.3016e+00 | -1.3016e+00 | -1.3016e+00 | -1.3016e+00 |
|  | Std | 2.4879e-11 | 6.7856e-10 | 6.1849e-16 | 1.8004e-08 | 4.7695e-11 | 6.6468e-16 | 8.5887e-09 | 1.3637e-10 |
| F17 | Ave | 3.9793e-01 | 3.9793e-01 | 3.9793e-01 | 3.9793e-01 | 3.9793e-01 | 3.9793e-01 | 3.9793e-01 | 3.9793e-01 |
|  | Std | 3.1854e-15 | 2.9802e-05 | 0.0000e+00 | 7.9053e-07 | 1.3227e-06 | 0.0000e+00 | 3.6674e-05 | 2.1220e-06 |
| F18 | Ave | 3.0000e+00 | 3.0000e+00 | 3.0000e+00 | 3.0000e+00 | 3.0000e+00 | 3.0000e+00 | 3.0000e+00 | 3.0000e+00 |
|  | Std | 1.0026e-02 | 1.4897e-15 | 1.4496e-15 | 1.7673e-05 | 9.4235e-09 | 9.7225e-16 | 4.9673e-06 | 8.5129e-06 |
| F19 | Ave | -3.8627e+00 | -3.8629e+00 | -3.8628e+00 | -3.8617e+00 | -3.8621e+00 | -3.8625e+00 | -3.8625e+00 | -3.8611e+00 |
|  | Std | 2.7952e-04 | 1.8903e-03 | 2.3557e-15 | 2.3308e-03 | 1.1021e-03 | 1.4489e-03 | 1.1065e-03 | 2.3873e-03 |
| F20 | Ave | -3.2703e+00 | -3.2907e+00 | -3.2576e+00 | -3.2655e+00 | -3.1252e+00 | -3.2534e+00 | -3.2294e+00 | -3.2106e+00 |

|     |     |           |            |            |            |            |            |            |            |            |
|-----|-----|-----------|------------|------------|------------|------------|------------|------------|------------|------------|
|     | Std | 6.0282e-02 | 4.5583e-02 | 5.7377e-02 | 6.9870e-02 | 1.2573e-01 | 7.2045e-02 | 1.1075e-01 | 7.9788e-02 |
| F21 | Ave | -7.2183e+00 | -1.0153e+01 | -8.9855e+00 | -9.6452e+00 | -5.2238e+00 | -8.0516e+00 | -1.0153e+01 | -9.2160e+00 |
|     | Std | 3.1258e+00 | 1.5309e-04 | 2.6852e+00 | 1.5462e+00 | 9.3088e-01 | 2.6529e+00 | 1.2311e-04 | 2.1513e+00 |
| F22 | Ave | -7.8623e+00 | -1.0402e+01 | -9.7499e+00 | -1.0226e+01 | -5.1847e+00 | -8.8053e+00 | -1.0402e+01 | -8.7848e+00 |
|     | Std | 3.2292e+00 | 1.9053e-03 | 2.0168e+00 | 9.6322e-01 | 1.0629e+00 | 2.7723e+00 | 2.4045e-04 | 2.7875e+00 |
| F23 | Ave | -6.3214e+00 | -1.0536e+01 | -1.0356e+01 | -1.0264e+01 | -5.6327e+00 | -9.4912e+00 | -1.0536e+01 | -8.6265e+00 |
|     | Std | 3.5528e+00 | 8.0933e-05 | 9.8730e-01 | 1.4813e+00 | 1.5467e+00 | 2.4760e+00 | 6.6312e-05 | 3.0473e+00 |

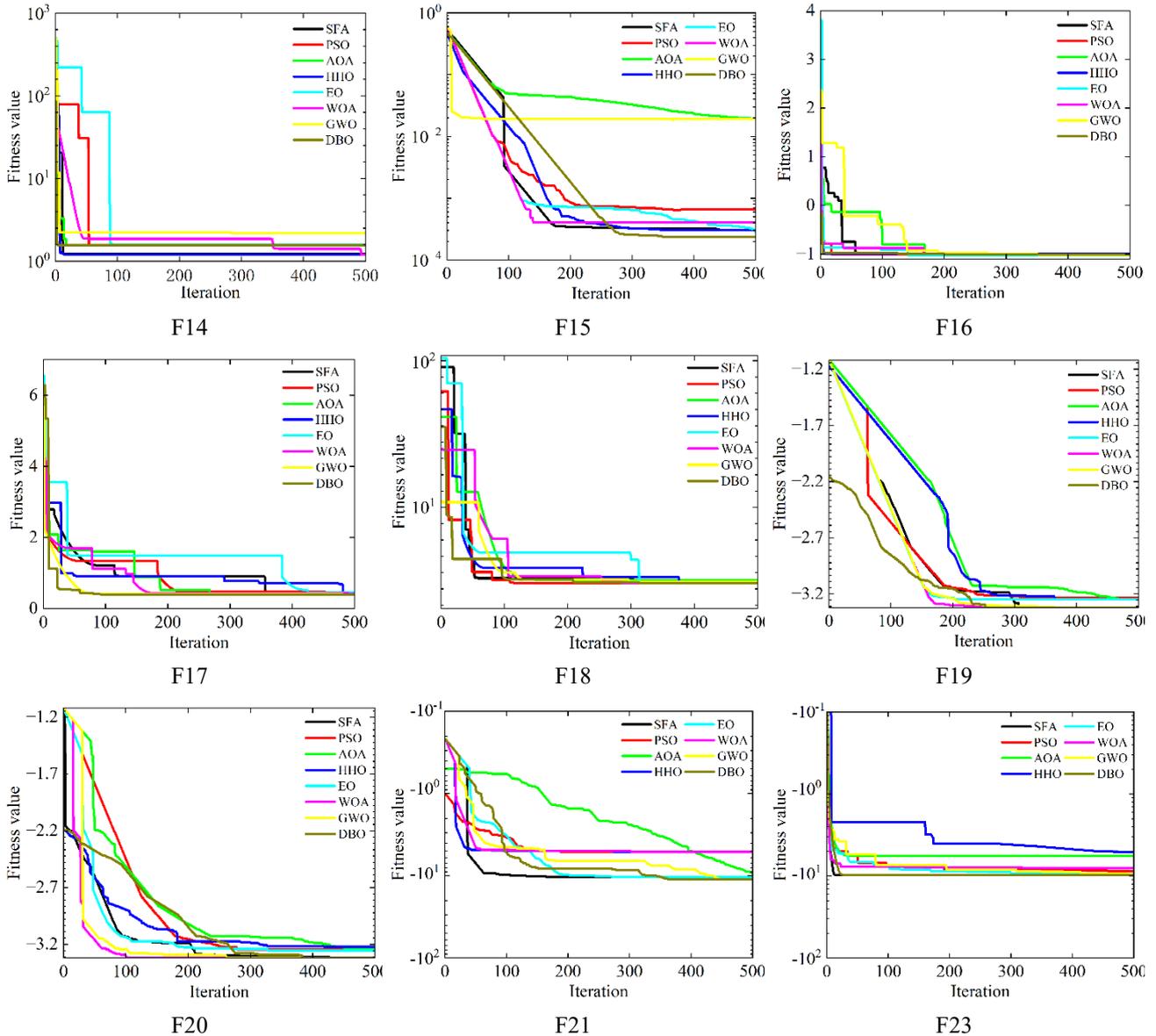

**Figure 12 Convergence curves of the algorithms for the fixed-dimension multimodal test functions.**

3.4 Evaluation on hybrid composition test functions (F24–F29)

Table 10 presents the results of hybrid composition functions (F24-F29). The experimental results demonstrate that DBO exhibits superior optimization performance in hybrid composition test functions (F24-F29). Figure 13 illustrates the convergence curves of the algorithms for the hybrid composition functions (F24-F29).

Table 10 Results of hybrid composition test functions (F24–F29).

|     |     | AOA | DBO | EO | GWO | HHO | PSO | SFA | WOA |
| --- | --- | --- | --- | --- | --- | --- | --- | --- | --- |
| F24 | Ave | 4.0712e+02 | 3.8318e+02 | 1.2993e+03 | 4.8735e+02 | 7.5213e+02 | 7.6839e+0.2 | 5.3687e+02 | 4.5137e+02 |
|     | Std | 8.1329e+01 | 3.6594e+01 | 1.5018e+02 | 1.4337e+02 | 5.7801e+01 | 7.6107e+01 | 1.0358e+02 | 2.0718e+02 |
| F25 | Ave | 9.1035e+02 | 9.0038e+02 | 1.0738e+03 | 9.8521e+02 | 9.5883e+02 | 1.1825e+03 | 9.3687e+02 | 9.0148e+02 |
|     | Std | 1.2124e+01 | 2.9074e+01 | 5.7868e+01 | 3.0003e+01 | 2.7745e+01 | 3.3017e+01 | 5.8412e+01 | 2.4719e+01 |
| F26 | Ave | 9.1204e+02 | 8.7705e+02 | 1.1078e+03 | 9.7420e+02 | 9.6216e+02 | 1.1809e+03 | 9.3953e+02 | 9.0410e+02 |
|     | Std | 3.9715e+00 | 1.9907e+01 | 4.0639e+01 | 2.2503e+01 | 4.5311e+01 | 3.5222e+01 | 4.2613e+01 | 3.9879e+01 |
| F27 | Ave | 9.1027e+02 | 9.0239e+02 | 1.1027e+03 | 9.7018e+02 | 9.6953e+02 | 1.2044e+03 | 9.3974e+02 | 9.0383e+02 |
|     | Std | 3.4325e+01 | 3.5048e+01 | 3.2677e+01 | 1.9547e+01 | 3.8227e+01 | 2.4037e+01 | 5.9879e+01 | 3.5513e+01 |
| F28 | Ave | 9.8135e+02 | 8.7507e+02 | 1.9657e+03 | 1.3429e+03 | 1.4778e+03 | 1.7138e+03 | 8.9638e+02 | 1.8762e+03 |
|     | Std | 4.7328e+01 | 3.9881e+01 | 1.2874e+01 | 1.9126e+02 | 2.4988e+02 | 3.5269e+01 | 5.8513e+01 | 2.3871e+02 |
| F29 | Ave | 1.6218e+03 | 1.5677e+03 | 2.1938e+03 | 1.9134e+03 | 1.6839e+03 | 2.1048e+03 | 1.6416e+03 | 1.5758e+03 |
|     | Std | 4.2527e+00 | 3.1923e+01 | 3.5578e+01 | 6.5738e+00 | 3.4900e+00 | 2.9726e+01 | 3.2213e+01 | 3.1682e+00 |

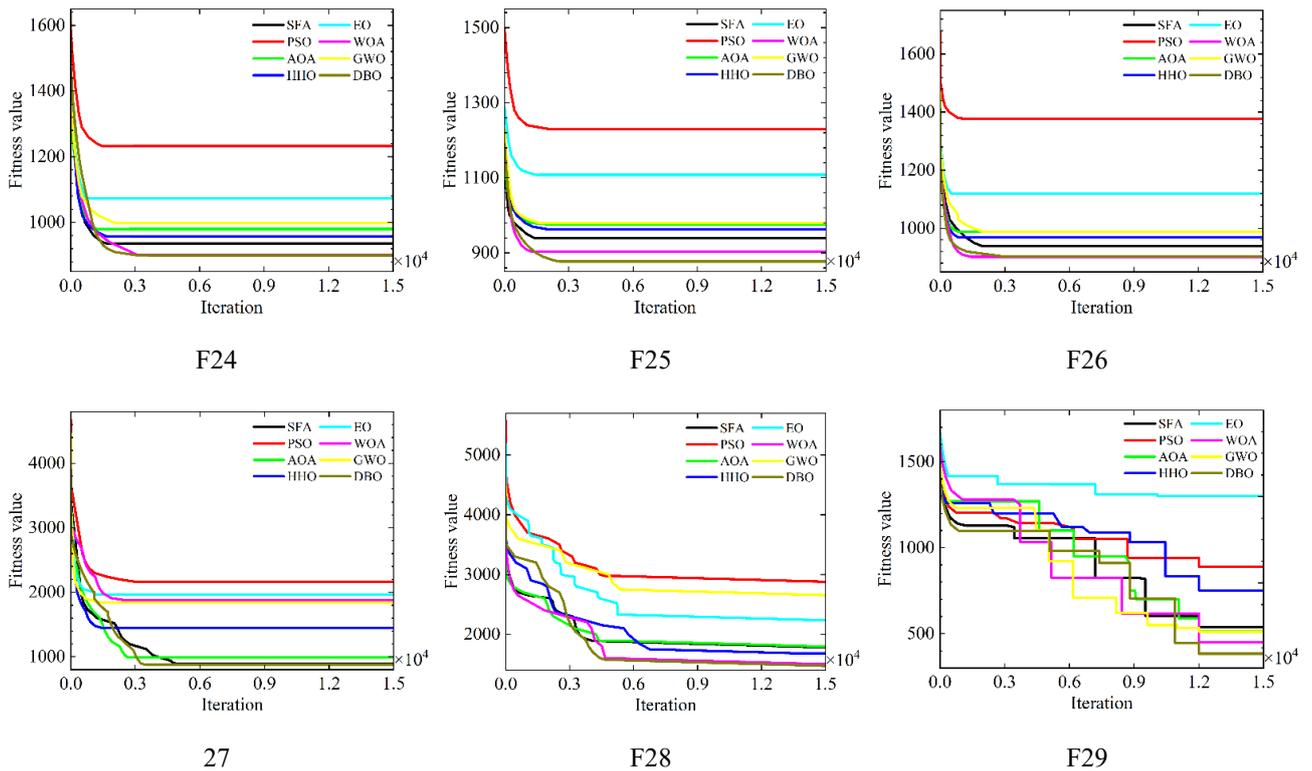

Figure 13 Convergence curves of the algorithms for the hybrid composition functions (F24-F29).

## 4. Classical Engineering Problems

In this section, we evaluate the optimization algorithms on two classical engineering problems: tension/compression spring design and pressure vessel design problem. Both problems are subject to design specifications, resource constraints, and other conditions that must be taken into account in the optimization process. The objective of constrained optimization problems is to find feasible solutions that accurately reflect the performance of the algorithms. To ensure consistency in the evaluation, the same experimental environment is employed for all algorithms: a total of 500 iterations and 30 search agents.

4.1 Tension/compression spring design problem

The objective of the tension/compression spring design problem is to minimize the weight of the spring while satisfying the constraints of minimum detection, shear stress, and surge frequency. Here, the constraint condition contains three variables: the diameter of the spring coil $E$, the diameter of the spring single wire $e$, and the number of coils in the spring $N$. The structure of tension/compression spring design problem is illustrated in Figure 14. Table 11 presents the optimization results for this problem.

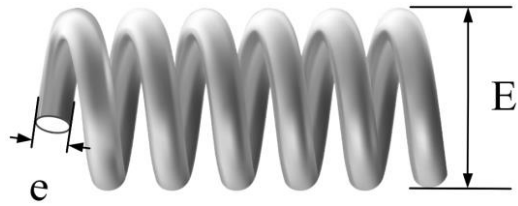

**Figure 14 Tension/compression spring design problem.**

Consider:

$$Consider: \vec{X} = [x_1\ x_2\ x_3] = [E\ e\ N]$$

$$Minimize: f(\vec{X}) = (x_3 + 2)x_2 x_1^2$$

$$Subject\ to: g_1(\vec{X}) = 1 - \frac{x_2^3 x_3}{71785 x_1^4} \leq 0$$

$$g_2(\vec{X}) = \frac{4x_2^2 - x_1 x_2}{12566(x_2 x_1^3 - x_1^4)} + \frac{1}{5108 x_1^2} - 1 \leq 0$$

$$g_3(\vec{X}) = 1 - \frac{140.45 x_1}{x_2^3 x_3} \leq 0$$

$$g_4(\vec{X}) = \frac{x_1 + x_2}{1.5} - 1 \leq 0$$

$$Variable\ range: 0.05 \leq x_1 \leq 2, 0.25 \leq x_2 \leq 1.3, 2 \leq x_3 \leq 15$$

**Table 11 Optimization results for tension/compression spring design problem.**

| Algorithm | Optimum variables | | | Optimum weight |
|---|---|---|---|---|
| | e | E | N | |
| BBO | 0.052808 | 0.383311 | 10.157006 | 0.0129951 |
| Constraint correction | 0.050000 | 0.315900 | 14.250000 | 0.0128334 |
| DE | 0.051609 | 0.354714 | 11.410831 | 0.0126702 |
| DBO | 0.051790 | 0.359148 | 11.147891 | 0.0126654 |
| ES | 0.051989 | 0.363965 | 10.890522 | 0.0126810 |
| GA | 0.051480 | 0.351661 | 11.632201 | 0.0127048 |
| GJO | 0.051580 | 0.354055 | 11.448400 | 0.0126675 |
| GSA | 0.050276 | 0.323680 | 13.525410 | 0.0127022 |
| GWO | 0.051690 | 0.356737 | 11.288850 | 0.0126660 |
| HHO | 0.051796 | 0.359305 | 11.138859 | 0.0126654 |
| Improved HS | 0.051154 | 0.349871 | 12.076432 | 0.0126706 |
| Mathematical optimization | 0.053396 | 0.399180 | 9.1854000 | 0.0127303 |
| PSO | 0.051728 | 0.357644 | 11.244543 | 0.0126747 |
| RFO | 0.051890 | 0.361420 | 11.584360 | 0.0132100 |
| RO | 0.051370 | 0.349096 | 11.762790 | 0.0126788 |
| SFA | 0.051651 | 0.355737 | 11.350000 | 0.0126697 |
| SSA | 0.051207 | 0.345215 | 12.004032 | 0.0126766 |
| WOA | 0.051200 | 0.345200 | 12.004000 | 0.0126725 |

The results of the experimental analysis indicate that DBO achieves the highest optimization performance and successfully minimizes the weight of the spring while adhering to the constraint conditions.

4.2 Pressure vessel design problem

The objective of the pressure vessel design problem is to minimize the production cost while ensuring safety. This problem is constrained by four variables: the inner radius $W$, the length of the shell $L$, the thickness of the head $T_h$ and the thickness of the vessel shell $T_s$. The structure of the pressure vessel design problem is illustrated in Figure 15. Table 12 presents the optimization results for this problem.

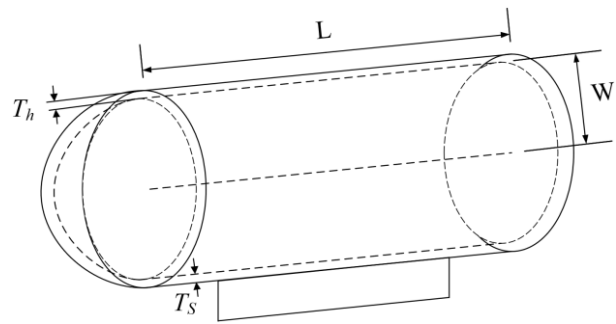

**Figure 15 Pressure vessel design problem.**

Consider:
$$Consider: \vec{X} = [x_1\ x_2\ x_3\ x_4] = [T_s\ T_h\ W\ L]$$

$$\text{Minimize: } f(\vec{X}) = 0.6224x_1x_3x_4 + 1.7781x_2x_3^2 + 3.1661x_1^2x_4 + 19.84x_1^2x_3$$
$$\text{Subject to: } g_1(\vec{X}) = -x_1 + 0.0193x_3 \leq 0$$
$$g_2(\vec{X}) = -x_2 + 0.00954x_3 \leq 0$$
$$g_3(\vec{X}) = -\pi x_3^2 x_4 - \frac{4}{3}\pi x_3^3 + 1296000 \leq 0$$
$$g_4(\vec{X}) = x_4 - 240 \leq 0$$
$$\text{Variable range: } 0 \leq x_1, x_2 \leq 99, 10 \leq x_3, x_4 \leq 200$$

**Table 12 Optimization results for pressure vessel design problem.**

| Algorithm | Optimum variables | | | | Optimum cost |
|---|---|---|---|---|---|
| | $T_s$ | $T_h$ | W | L | |
| ACO | 0.812500 | 0.437500 | 42.103624 | 176.572656 | 6059.0888 |
| AOA | 0.830374 | 0.416206 | 42.751270 | 169.345400 | 6048.7844 |
| BBO | 1.333988 | 0.095568 | 62.187134 | 69.4430941 | 6181.8327 |
| Branch-bound | 1.125000 | 0.625000 | 47.700000 | 117.701000 | 8129.1036 |
| DE | 0.812500 | 0.437500 | 42.098411 | 176.637690 | 6059.7340 |
| DBO | 0.823574 | 0.409206 | 42.651270 | 169.915400 | 5977.2759 |
| ES | 0.812500 | 0.437500 | 42.098087 | 176.640518 | 6059.7456 |
| GA | 0.812500 | 0.434500 | 40.323900 | 200.000000 | 6288.7445 |
| GSA | 1.125000 | 0.625000 | 55.988660 | 84.4542025 | 8538.8359 |
| GWO | 0.812500 | 0.434500 | 42.089181 | 176.758731 | 6051.5639 |
| HHO | 0.851200 | 0.422400 | 44.088650 | 153.545600 | 6032.6746 |
| Improved HS | 1.125000 | 0.625000 | 58.290150 | 43.6926800 | 7197.7300 |
| Lagrangian multiplier | 1.125000 | 0.625000 | 58.291000 | 43.6900000 | 7198.0428 |
| PSO | 0.812500 | 0.437500 | 42.091266 | 176.746500 | 6061.0777 |
| RFO | 0.814250 | 0.445210 | 42.2023100 | 176.621450 | 6113.3195 |
| RSA | 0.840100 | 0.419000 | 43.3812000 | 161.555600 | 6034.7591 |
| WOA | 0.812500 | 0.437500 | 42.0982699 | 176.638998 | 6059.7410 |

The results of the experimental analysis indicate that DBO achieves the highest optimization performance under the condition of satisfying constraints, and minimizes the production cost of pressure vessel under the premise of ensuring safety.

5. **Conclusion**

This paper proposes a novel human-based meta-heuristic named dragon boat optimization, inspired by the dragon boat racing. The algorithm is meticulously designed by modeling the behavior patterns of a dragon boat crew during racing events. Firstly, the crew is categorized into three roles: drummer, paddler, and steersman, with the incorporation of social psychology mechanisms to capture real-world behavior accurately. Secondly, the collaborative relationship between the drummer and paddlers is analyzed, and the acceleration factor and attenuation factor are introduced to dynamically update the states of the paddlers. Furthermore, the imbalance rate is introduced to evaluate the impact of varying water surface conditions on paddlers' performance. Finally, in each iteration, the state update strategy of a dragon boat is modeled, leading to the dragon boat optimization algorithm. A comparative analysis of the test results between our algorithm and other optimization algorithms for 29 benchmark functions and two classical engineering problems shows that the dragon boat optimization algorithm achieves state-of-the-art performance while maintaining low computational complexity.

**Acknowledgement**

This work is supported by the China Scholarship Council; the National Natural Science Foundation of China: 91948303-1; the National Key R&D Program of China (2021ZD0140301); the National Natural Science Foundation of China: 611803375，No. 61803375，No. 12002380, No. 62106278, No. 62101575，No. 61906210; the Postgraduate Scientific Research Innovation Project of Hunan Province: QL20210018.

**Appendix**

In this section, we present the 3-D maps and convergence curves for F6 and F22. They are shown in Figure 16.

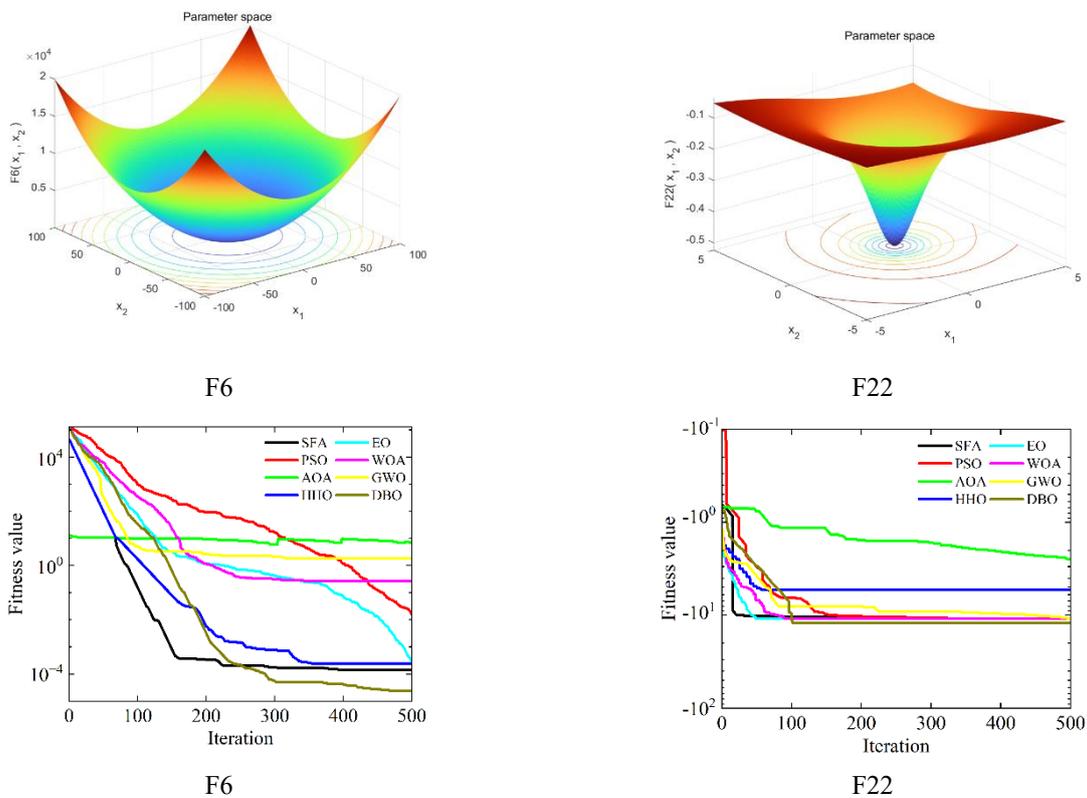

**Figure 16 3-D maps and convergence curves of F6 and F22.**

**References**

[1]  K. Rajwar, K. Deep, and S. Das, "An exhaustive review of the metaheuristic algorithms for search and optimization: taxonomy, applications, and open challenges," (in English), *ARTIFICIAL INTELLIGENCE REVIEW,* 2023 APR 9 2023, doi: 10.1007/s10462-023-10470-y.

[2]  M. Kaveh and M. S. Mesgari, "Application of Meta-Heuristic Algorithms for Training Neural Networks and Deep Learning Architectures: A Comprehensive Review," (in English), *NEURAL PROCESSING LETTERS,* vol. 55, no. 4, pp. 4519-4622, AUG 2023, doi: 10.1007/s11063-022-11055-6.

[3]  S. Darvishpoor, A. Darvishpour, M. Escarcega, and M. Hassanalian, "Nature-Inspired Algorithms from Oceans to Space: A Comprehensive Review of Heuristic and Meta-Heuristic Optimization Algorithms and Their Potential Applications in Drones," (in English), *DRONES,* vol. 7, no. 7, JUL 2023, Art no. 427, doi: 10.3390/drones7070427.

[4]  O. W. Khalid, N. A. M. Isa, and H. A. M. Sakim, "Emperor penguin optimizer: A comprehensive review based on state-of-the-art meta-heuristic algorithms," (in English), *ALEXANDRIA ENGINEERING JOURNAL,* vol. 63, pp. 487-526, FEB 2023, doi: 10.1016/j.aej.2022.08.013.

[5]  P. Moscato, "On Evolution, Search, Optimization, Genetic Algorithms and Martial Arts - Towards Memetic Algorithms," 2000.

[6]  S. Das and P. N. Suganthan, "Differential Evolution: A Survey of the State-of-the-Art," *IEEE Transactions on Evolutionary Computation,* vol. 15, no. 1, pp. 4-31, 2011, doi: 10.1109/TEVC.2010.2059031.

[7]  I. Rahman and J. Mohamad-Saleh, "Hybrid bio-Inspired computational intelligence techniques for solving power system optimization problems: A comprehensive survey," *Applied Soft Computing,* vol. 69, pp. 72-130, 2018/08/01/ 2018, doi: https://doi.org/10.1016/j.asoc.2018.04.051.

[8]  C. Ryan, J. J. Collins, and M. O. Neill, "Grammatical evolution: Evolving programs for an arbitrary language," in *Genetic Programming*, Berlin, Heidelberg, W. Banzhaf, R. Poli, M. Schoenauer, and T. C. Fogarty, Eds., 1998// 1998: Springer Berlin Heidelberg, pp. 83-96.

[9]  Koza and JohnR, *Genetic programming : on the programming of computers by means of natural selection*. Genetic programming : on the programming of computers by means of natural selection, 1992.

[10] X.-S. Yang, "A new metaheuristic bat-inspired algorithm," in *Nature inspired cooperative strategies for optimization (NICSO 2010)*: Springer, 2010, pp. 65-74.

[11] D. Teodorović, "Bee colony optimization (BCO)," in *Innovations in swarm intelligence*: Springer, 2009, pp. 39-60.

[12] K. C. O. Roeva, "Cuckoo search and firefly algorithms in terms of generalized net theory," *Soft computing: A fusion of foundations, methodologies and applications,* vol. 24, no. 7, 2020.

[13] D. Karaboga, "An idea based on honey bee swarm for numerical optimization," Technical report-tr06, Erciyes university, engineering faculty, computer …, 2005.

[14] M. Dorigo, V. Maniezzo, and A. Colorni, "Ant system: optimization by a colony of cooperating agents," *IEEE transactions on systems, man, and cybernetics, part b (cybernetics),* vol. 26, no. 1, pp. 29-41, 1996.

[15] S. Mirjalili, S. M. Mirjalili, and A. Lewis, "Grey Wolf Optimizer," *Advances in Engineering Software,* vol. 69, pp. 46-61, 2014/03/01/ 2014, doi: https://doi.org/10.1016/j.advengsoft.2013.12.007.

[16] S. Mirjalili and A. Lewis, "The Whale Optimization Algorithm," *Advances in Engineering Software,* vol. 95, pp. 51-67, 2016/05/01/ 2016, doi: https://doi.org/10.1016/j.advengsoft.2016.01.008.

[17] M. D. Li, H. Zhao, X. W. Weng, and T. Han, "A novel nature-inspired algorithm for optimization: Virus colony search," *Advances in Engineering Software,* vol. 92, pp. 65-88, 2016/02/01/ 2016, doi: https://doi.org/10.1016/j.advengsoft.2015.11.004.

[18] A. H. Gandomi and A. H. Alavi, "Krill herd: a new bio-inspired optimization algorithm," *Communications in nonlinear science and numerical simulation,* vol. 17, no. 12, pp. 4831-4845, 2012.

[19] O. K. Erol and I. Eksin, "A new optimization method: Big Bang–Big Crunch," *Advances in Engineering Software,* vol. 37, no. 2, pp. 106-111, 2006/02/01/ 2006, doi: https://doi.org/10.1016/j.advengsoft.2005.04.005.

[20] E. Rashedi, H. Nezamabadi-pour, and S. Saryazdi, "GSA: A Gravitational Search Algorithm," *Information Sciences,* vol. 179, no. 13, pp. 2232-2248, 2009/06/13/ 2009, doi: https://doi.org/10.1016/j.ins.2009.03.004.

[21] E. Cuevas, D. Oliva, D. Zaldivar, M. Perez-Cisneros, and H. Sossa, "Circle detection using electro-magnetism optimization," 2014.

[22] Z. W. Geem, J. H. Kim, and G. V. Loganathan, "A new heuristic optimization algorithm: harmony search," *simulation,* vol. 76, no. 2, pp. 60-68, 2001.

[23] Z. Wei, C. Huang, X. Wang, T. Han, and Y. Li, "Nuclear Reaction Optimization: A Novel and Powerful Physics-Based Algorithm for Global Optimization," *IEEE Access,* vol. 7, pp. 66084-66109, 2019, doi: 10.1109/ACCESS.2019.2918406.

[24] S. Kirkpatrick, C. D. Gelatt, and M. P. Vecchi, "Optimization by Simulated Annealing," in *Readings in Computer Vision*, M. A. Fischler and O. Firschein Eds. San Francisco (CA): Morgan Kaufmann, 1987, pp. 606-615.

[25] S. Mirjalili, "SCA: A Sine Cosine Algorithm for solving optimization problems," *Knowledge-Based Systems,* vol. 96, pp. 120-133, 2016/03/15/ 2016, doi: https://doi.org/10.1016/j.knosys.2015.12.022.

[26] E. Fadakar and M. Ebrahimi, "A new metaheuristic football game inspired algorithm," in *2016 1st Conference on Swarm Intelligence and Evolutionary Computation (CSIEC)*, 9-11 March 2016 2016, pp. 6-11, doi: 10.1109/CSIEC.2016.7482120.

[27] A. Kaveh and M. Kooshkebaghi, "artificial coronary circulation system: a new bio-inspired metaheuristic algorithm," *Scientia Iranica,* 2019.

[28] O. E. Turgut, M. Asker, and M. T. Oban, "Artificial Cooperative Search Algorithm for Optimal Loading Of Multi-Chiller Systems," 2015.

[29] J. D. Farmer, N. H. Packard, and A. S. Perelson, "The immune system, adaptation, and machine learning," *Physica D: Nonlinear Phenomena,* vol. 22, no. 1, pp. 187-204, 1986/10/01/ 1986, doi: https://doi.org/10.1016/0167-2789(86)90240-X.

[30] L. N. d. Castro and J. Timmis, "An artificial immune network for multimodal function optimization," in *Proceedings of the 2002 Congress on Evolutionary Computation. CEC'02 (Cat. No.02TH8600)*, 12-17 May 2002 2002, vol. 1, pp. 699-704 vol.1, doi: 10.1109/CEC.2002.1007011.

[31] N. Ghorbani and E. Babaei, "Exchange market algorithm," *Applied Soft Computing,* vol. 19, pp. 177-187, 2014/06/01/ 2014, doi: https://doi.org/10.1016/j.asoc.2014.02.006.

[32] P. Savsani and V. Savsani, "Passing vehicle search (PVS): A novel metaheuristic algorithm," *Applied Mathematical Modelling,* vol. 40, no. 5, pp. 3951-3978, 2016/03/01/ 2016, doi: https://doi.org/10.1016/j.apm.2015.10.040.

[33] Y. Xu, Z. Cui, and J. Zeng, "Social Emotional Optimization Algorithm for Nonlinear Constrained Optimization Problems," in *Swarm, Evolutionary, and Memetic Computing*, Berlin, Heidelberg, B. K. Panigrahi, S. Das, P. N. Suganthan, and S. S. Dash, Eds., 2010// 2010: Springer Berlin Heidelberg, pp. 583-590.

[34] O. A. Raouf *et al.*, "Sperm motility algorithm: a novel metaheuristic approach for global optimisation articial coronary circulation system: a new bio-inspired metaheuristic algorithm Artificial Cooperative Search Algorithm for Optimal Loading Of Multi-Chiller Systems," *International Journal of Operational Research,* vol. 28, no. 2, 2017.